\documentclass[10pt,letter,aps,prl,nofootinbib]{revtex4}

\usepackage{graphicx}
\usepackage{epstopdf}
\usepackage{amssymb}

\begin{document}
\title{Local electron and ionic heating effects on
the conductance of nanostructures}
\author{Roberto D'Agosta and Massimiliano Di Ventra}
\address{Department of Physics, University of California - San Diego,
La Jolla, CA 92093-0319}
\date{\today}

\begin{abstract} Heat production and dissipation
induced by current flow in nanostructures is of primary importance
to understand the stability of these systems. These effects have
contributions from both electron-phonon and electron-electron
interactions. Here, we consider the effect of the local electron and
ionic heating on the conductance of nanoscale systems. Specifically
we show that the non-linear dependence of the conductance on the
external bias may be used to infer information about the local
heating of both electrons and ions. We compare our results with
available experimental data on transport in $\mathrm{D}_2$ and $\mathrm{H}_2$
molecules. The comparison between experiment and theory is
reasonably good close to the lowest phonon mode of the molecule,
especially for the $\mathrm{D}_2$ molecule. At higher biases we cannot rule
out the presence of other effects like, e.g., current-induced forces
that make the scenario more complex.
\end{abstract}

\maketitle

\section{Introduction}
The idea of building electronic devices from nanostructures has
gathered a lot of attention due to the high expectations in terms of
size reduction and power dissipation \cite{DiVentra2004}.
Encouraging progress has been made in experimental techniques and
theoretical modeling towards this aim \cite{DiVentra2008}. However,
a fundamental and technologically important issue, namely local heat
production and dissipation in these systems has attracted much less
attention \cite{Todorov1998,Segal2002,Chen2003,Chen2005,
DAgosta2006c,Zhifeng2006,Zhifeng2007,Pecchia2007}.

It has been argued that since the electron inelastic mean free path
is large compared to the dimensions of a nanostructure, no energy
dissipation occurs in the nanostructure region. However, nanoscale
systems carry very large current densities compared to bulk
electrodes. This implies an increased number of scattering events
per unit time and unit volume so that interactions among electrons
or among electrons and phonons are particularly important. In
addition, the reduced size means a small heat capacitance: any small
energy transfer from the current-carrying electrons to local ionic
vibrations or other electrons in the system may induce a substantial
heating of the nanostructure \cite{DAgosta2006c}.

So far, direct measurements of the amount of energy locally
dissipated in a nanoscale system have been beyond our reach.
However, new experiments have considered the indirect effects of
local heating on accessible quantities
\cite{Zhifeng2006,Zhifeng2007,Tsutsui2007,Tsutsui2006}. For example,
in \cite{Zhifeng2006,Zhifeng2007} an effective ionic temperature is
determined via the force needed to break the chemical bonds between
molecules and the adjacent leads. These experiments indirectly probe
the local ionic temperature, the contributions due to
electron-electron interactions, and corresponding local electron
heating \cite{DAgosta2006c,Zhifeng2007}.

Here, we discuss another possible indirect method to probe both the
local ionic and electron temperatures via the non-linearities in the
DC conductance of nanostructures. We will compare our results with
the experimental conductance of simple molecules such as
$\mathrm{D_2}$ and $\mathrm{H_2}$ sandwiched between two $\mathrm{Pt}$
leads as studied in Ref. \cite{Djukic2005} (and references therein).

In order to address the above issues we need a theory that takes
into account both energy production and dissipation on an equal
footing. A full quantum-mechanical description in terms of many-body
states for the present non-equilibrium problem seems hopeless.
Instead, we have previously shown that a much more ``economical''
hydrodynamic theory in terms of the single-particle density and
current density may be derived for nanostructures
\cite{DAgosta2006a}. In this paper, we first review such a
theory, and later on use it to study the effect of heating on
conductance.

\section{Classical hydrodynamics}

In the following, we will refer to some concepts of classical
hydrodynamics. For completeness, we repeat here some of those
concepts, while a more comprehensive description of the dynamics of
classical fluids can be found in many textbooks \cite{Landau6,
Goldstein1965}.

The dynamics of a classical viscous fluid is usually described by
the so-called Navier-Stokes equations for the single-particle
density, $n(r,t)$, and the velocity field, $v(r,t)$, (ratio between
the current density and the density)
\begin{eqnarray}
D_t n(r,t)=-n(r,t)\nabla\cdot v(r,t)\nonumber\\
mn(r,t)D_t v_i(r,t)=-\nabla_i P(r,t)+\nabla_j
\pi_{i,j}(r,t)-n(r,t)\nabla_i V_{ext}(r,t)
\label{navier-stokes}
\end{eqnarray}
where $P(r,t)$ is the pressure, $\pi_{i,j}(r,t)$ is
the {\it Navier-Stokes stress tensor}
\begin{eqnarray}
\pi_{i,j}(r,t)&=&\eta \left[\nabla_j
v_i(r,t)+\nabla_i v_j(r,t)-\frac23 \delta_{i,j}\nabla\cdot
v(r,t)\right ]\nonumber\\ &&+\zeta \delta_{i,j}\nabla\cdot v(r,t),
\end{eqnarray}
and $V_{ext}(r,t)$ is the external potential. [Throughout the paper,
$\nabla_i$ corresponds to the derivative with respect to the $i$-th
spatial component ($i=\{x,y,z\}$), and summation over repeated
indexes is understood.] In these equations, the operator
$D_t=\partial_t+v(r,t)\cdot\nabla$ is the so-called ``convective"
derivative, while the viscosity coefficients, $\eta$ and $\zeta$,
are the shear and bulk viscosity of the liquid, respectively. The
viscosity coefficients have their origin in the approximate nature
of the Navier-Stokes equations and in the particle-particle
interaction \cite{Ghosh1982}. The first equation in
(\ref{navier-stokes}) is the continuity equation and states the mass
conservation when sources or sinks are not present. The second
equation in (\ref{navier-stokes}) is the force equation: the left
hand side is the acceleration of a small volume of liquid subject to
the internal forces (due to pressure, particle-particle
interactions) and external forces ($F_{ext}=-\nabla V_{ext}$). It is
important to realize the approximate nature of these equations: in
classical physics the very basic concept of particle density $n$ has
a meaning only in a coarse-grained sense, i.e., with respect to
volumes of the liquid small compared to the other relevant scales of
the problem, but large to contain ``enough'' particles so that a
continuum mechanics can be developed. In the opposite condition, one
has to revert to the solution of the Newton equations of motion for
each particle. Due to the continuous spatial nature of
wave-functions, the above limitations do not pertain to Quantum
Mechanics, for which a hydrodynamic description can be formulated
exactly.

\section{Hydrodynamical formulation of Quantum Mechanics}

Ever since the formulation of the Schr\"odinger equation of motion
for complex wave-functions, there have been several attempts to
formulate Quantum Mechanics in terms of classical quantities. The
degree to which these attempts have been successful is still
undecided, since the use of words like ``particle'',
``trajectories'', and ``directions of propagation" is widespread in
the modern scientific literature. One such attempt was made at the
dawn of Quantum Mechanics, in 1926, by Madelung
\cite{Madelung1926,Ghosh1982} who showed that the Schr\"odinger
equation for a single particle is exactly equivalent to a set of
equations of motion for the particle density and ``velocity''. For
this single-particle problem, the velocity is defined as the
variation of the phase of the wave-function with position, and thus
seems a mere mathematical tool \cite{Ghosh1982}. An equivalent, but
more transparent definition of this velocity field is
\begin{equation}
v(r,t)=\frac{j(r,t)}{n(r,t)},
\end{equation}
where $j(r,t)$ is the current density.
This definition is valid for the points $r$ for which $n(r,t)\not =0$.
It is remarkable that the equation of motion for this velocity is
governed by the external forces, plus a ``quantum mechanical"
contribution, known as ``Bohm stress tensor", that has not a classical
counterpart \cite{Ghosh1982}. Indeed, if we start from the Schr\"odinger equation for
the wave-function, $\Psi$, of a particle in the presence of the
external potential $V_{ext}$, ($\hbar=e=1$ throughout this paper, where $e$ is the electron charge)
\begin{equation}
i \partial_t \Psi(r,t)=-\frac{1}{2m}\nabla^2
\Psi(r,t)+V_{ext}(r,t)\Psi(r,t)
\label{schroedinger}
\end{equation}
we can rewrite $\Psi(r,t)$ in terms of two real
functions of time and position, $R$ and $S$, as
\begin{equation}
\Psi(r,t)=R(r,t)e^{iS(r,t)}.
\end{equation}
It is a simple exercise to show that, if one {\it
defines} the density $n(r,t)$, and the velocity $v(r,t)$,
\begin{eqnarray}
n(r,t)=R^2(r,t)=|\Psi(r,t)|^2,\\
v(r,t)=\frac{\nabla S(r,t)}{m},
\end{eqnarray}
then the equations of motion
\begin{eqnarray}
\partial_t n(r,t)=-\nabla
\cdot\left[n(r,t)v(r,t)\right],
\label{continuity}\\
m\partial_t v(r,t)=\frac{1}{2m}\nabla \left(\frac{\nabla^2
R(r,t)}{R(r,t)}\right)-m v(r,t)\cdot \nabla v(r,t)-\nabla V_{ext}(r,t)
\label{newton}
\end{eqnarray}
hold. Eqs.~(\ref{continuity}) and (\ref{newton}) have a
clear physical interpretation: The quantum mechanical system is
equivalent to a fluid whose dynamics is governed by the Euler equation (\ref{newton})
subjected to the force exerted by the external potential
\cite{Landau6}, and an internal force whose origin is purely quantum
mechanical. \footnote{The first term on the right hand side of
Eq.~(\ref{newton}).} Moreover, the dynamics conserves the {\it mass},
i.e., the total probability, and then the continuity equation
(\ref{continuity}) holds \cite{Sakurai}. The solution of the equations
of motion (\ref{continuity}) and (\ref{newton}) {\it is equivalent} to
the solution of the Schr\"odinger equation. It is interesting to point out that the quantum
mechanical force can be expressed in terms of the Bohm {\it stress tensor},
\begin{equation}
P_{i,j}\equiv-\frac{1}{4m}n(r,t)\nabla_i \nabla_j
\ln(n(r,t))
\end{equation}
and
\begin{equation}
\frac{n(r,t)}{2m}\nabla_i \left(\frac{\nabla^2
R(r,t)}{R(r,t)}\right)=-\nabla_j P_{i,j}(r,t).
\end{equation}
If one introduces the convective derivative the equations of motion
(\ref{continuity}) and (\ref{newton}) assume the well known form of
the {\it Navier-Stokes} equations of motion
\begin{eqnarray}
D_tn(r,t)=-n(r,t)\nabla\cdot v(r,t),
\label{ns-continuity}\\ 
mn(r,t)D_t v_i(r,t)=-\nabla_j
P_{i,j}(r,t)-n(r,t)\nabla_i V_{ext}(r,t).
\label{ns-newton}
\end{eqnarray}
Eqs.~(\ref{ns-continuity}) and (\ref{ns-newton}) are formally
identical to the Navier-Stokes equations~(\ref{navier-stokes}) for a
classical fluid. However, unlike the Navier-Stokes equations which
describe an approximate dynamics of the many-body classical fluid,
Eqs.~(\ref{ns-continuity}) and (\ref{ns-newton}) are {\em exactly}
equivalent to the Schr\"odinger equation: no approximation has been
made in their derivation.

While this approach to Quantum Mechanics may appear as a simple
attempt to recover a classical description of quantum phenomena,
over the years it has proven to be a very useful tool to describe
the dynamics of quantum systems in several contexts ranging from
condensed matter physics to nuclear physics (see, e.g.,
\cite{Kan1977}, and references therein). More recently, we have
shown that a hydrodynamic description of the electron flow in
nanoscale systems leads to the prediction of novel phenomena, like
the existence of a dynamical (viscous) resistance \cite{Sai2005},
turbulence \cite{DAgosta2006a,Sai2007,Bushong2007a,Bushong2007b},
and local electron heating and its effect on ionic heating
\cite{DAgosta2006c,Zhifeng2007}.

Here, we describe our hydrodynamical approach to transport in
nanostructures. As a first step, we need to generalize the
derivation of the equations of motion
(\ref{ns-continuity})-(\ref{ns-newton}) to the case of a many-body
interacting system. We follow closely the formalism presented in
Refs.~\cite{DAgosta2006a,Tokatly2005a}. (See also Ref.
\cite{Martin1959} for a general formulation of the dynamics of a
many-particle electron system.) We describe the dynamics of the
system via a field creation (annihilation) operator
$\psi^\dagger(r,t)$ ($\psi(r,t)$) which evolves in time following
the Heisenberg equation of motion
\begin{eqnarray} i\partial_t
\psi(r,t)&=&-\frac{1}{2m}\nabla^2\psi(r,t)+V_{ext}(r,t)\psi(r,t)\nonumber
\\ &&+\int dr'~\psi^\dagger(r',t)w(|r-r'|)\psi(r',t)\psi(r,t),
\end{eqnarray}
where the potential $w(|r-r'|)$ describes the particle-particle
interaction.  We define the single-particle density operator via the
usual definition, $\hat n(r,t)=\psi^\dagger(r,t)\psi(r,t)$ and the
current density operator
\begin{equation} \hat j(r,t)=\frac{i}{2m}\left[ \left(\nabla
\psi^\dagger(r,t)\right)\psi(r,t)-\psi^\dagger(r,t)\nabla\psi(r,t)
\right].
\end{equation}
It is lengthy but straightforward to show that these
two operators follow the dynamics induced by the coupled equations of
motion
\begin{eqnarray}
\partial_t \hat n(r,t)&=&-\nabla\cdot \hat j(r,t)\\
m\partial_t \hat j_i(r,t)&=&-\hat n(r,t)\nabla_i V_{ext}(r,t)-\nabla_j \hat T_{i,j}(r,t)\nonumber\\
&&-\psi^\dagger(r,t)\int dr'~\psi^\dagger(r',t)\nabla_i
w(|r-r'|)\psi(r',t)\psi(r,t)
\end{eqnarray}
where we have defined the kinetic stress tensor operator
\begin{eqnarray}
\hat T_{i,j}(r,t)&=&\frac{1}{2m}\left[\nabla_i
\psi^\dagger(r,t)\nabla_j \psi(r,t)+\nabla_j \psi^\dagger(r,t)\nabla_i
\psi(r,t)\right.\nonumber\\
&&\left.-\frac{\delta_{i,j}}{2} \nabla^2 \hat n(r,t)\right].
\end{eqnarray}
From the equations of motion for the operators, we get
immediately the equations of motion for their expectation values
\begin{eqnarray}
\partial_t n(r,t)&=&-\nabla\cdot j(r,t) \label{continuity-mb}\\
m\partial_t j_i(r,t)&=&-n(r,t)\nabla_i V_{ext}(r,t)-\nabla_j\langle
\hat T_{i,j}(r,t)\rangle \nonumber\\ &&-\int dr'~\rho_2(r,r',t)\nabla_i
w(|r-r'|)\label{secondeq}
\end{eqnarray}
where $\rho_2(r,r',t)=\langle
\psi^\dagger(r,t)\psi^\dagger(r',t)\psi(r',t)\psi(r,t)\rangle$.
Another rather lengthy and involved calculation allows us to write the
force density due to the particle-particle interaction as a
second-rank tensor, provided the interactions are negligibly
small at the boundary of the integration volume in equation~(\ref{secondeq}).
The result is \cite{DAgosta2006a,Tokatly2005a}
\begin{eqnarray}
\nabla_j W_{i,j}&=&-\frac12
\nabla_j\int dy~\frac{y_iy_j}{|y|} \frac{dw(|y|)}{d|y|}\int_0^1
d\lambda~\rho_2(r+\lambda y,r-(1-\lambda)y,t)\nonumber\\
&\equiv&\int
dr'~\rho_2(r,r',t)\nabla_i w(|r-r'|),
\end{eqnarray}
so that we arrive at the dynamical equation
\begin{equation}
m\partial_t j(r,t)=-n(r,t)\nabla_i
V_{ext}(r,t)-\nabla_j P_{i,j}
\label{newton-mb}
\end{equation}
where we have defined
\begin{equation} P_{i,j}=W_{i,j}+\langle \hat T_{i,j}\rangle.
\label{total-stress}
\end{equation}
From here, by using the definition of convective derivative and
re-scaling the particle momentum so that the stress tensor reads
\begin{equation}
P_{i,j}=W_{i,j}+\langle
\hat T_{i,j}\rangle-m\,n\,v_i\,v_j,
\end{equation}
one obtains the equations of motion for the particle and current
densities in a form identical to the single-particle equations of
motion (\ref{ns-continuity})-(\ref{ns-newton}).

Like Eqs.~(\ref{ns-continuity}) and (\ref{ns-newton}), which, for a
given initial condition, constitute a {\it closed} set, i.e., their
solution is equivalent to the solution of the single-particle
time-dependent Schr\"odinger equation, also their many-body
counterpart, Eqs.~(\ref{continuity-mb}) and~(\ref{newton-mb}) are
{\it equivalent} to the solution of the many-body time-dependent
Schr\"odinger equation. This equivalence is a direct consequence of
the theorems of time-dependent density-functional theory
\cite{Marques,Runge1984}. These theorems state that, given an
initial condition, there exists a one-to-one correspondence between
the time evolution of the particle density $n(r,t)$ and scalar
potential $V_{ext}(r,t)$ applied to the quantum mechanical system. A
similar correspondence holds between the current density $j(r,t)$
and an external {\it vector} potential $A(r,t)$
\cite{Marques,Ghosh1988,Vignale1996,DiVentra2007}, while the mapping
does not generally exist between the current density and the
external scalar potential \cite{DAgosta2005a}. The physical
relevance of these theorems to our case is that the stress tensor
$P_{i,j}(r,t)$ in (\ref{total-stress}) is a functional of either the
density or the current density, i.e.,
$P_{i,j}(r,t)=P_{i,j}[n(r,t'),t]$ or
$P_{i,j}(r,t)=P_{i,j}[j(r,t'),t]$ (with $t'\leq t$). This implies
that once the exact many-body stress tensor $P_{i,j}$ is known, one
can, in principle, recover from the solution of
Eqs.~(\ref{continuity-mb}) and (\ref{newton-mb}) full information on
the many-body wave-function.

Needless to say, the exact stress tensor is unknown. However,
starting from Eqs.~(\ref{continuity-mb}) and (\ref{newton-mb}) one
can develop perturbation schemes to approximate the exact stress
tensor~\cite{Vignale1996,Vignale1996b,Vignale1997b}, at least for
the problem at hand, thus simplifying enormously the solution of the
many-body problem. In the following, we will describe one of such
approximation schemes for the present case of current flow in a
nanojunction. We will derive an equation of motion for the stress
tensor $P_{i,j}$ and show that it depends on the so-called
three-particle stress tensor $P^{(3)}_{i,j,k}$, which in turn
describes the way three particles interact. The derivation of an
equation of motion for $P^{(3)}_{i,j,k}$ would bring us into the maze of
a hierarchic set of equations for stress tensors that describe
electron-electron interactions to all orders. We will show, however,
that for the case at hand, we can truncate this hierarchy and obtain
a closed equation for the stress tensor $P_{i,j}$.

\section{Visco-elasticity of the electron liquid} In parallel with the
hydrodynamic description of Quantum Mechanics, a visco-elastic
formulation of the dynamics of the electron liquid has been derived
within linear-response theory. It has been realized that a certain
class of low-energy, long-wavelength excitations of the electron
liquid may be mapped into the dynamics of a {\it visco-elastic}
medium \cite{Conti1999}. The dynamics of this visco-elastic medium
is described by an equation of motion for the current density given
by (in linear response and $d$ dimensions, $d>1$)
\begin{eqnarray}
m n\partial_t v(r,t)&=&\left[ \tilde
K+\left(1-\frac{2}{d}\right)\tilde
\mu\right]\nabla\left(\nabla\cdot v(r,t)\right)\nonumber\\
&&+\tilde \eta \nabla^2 v(r,t)-n(r,t)\nabla V_{ext}(r,t)
\label{visco-elastic}
\end{eqnarray}
where $\tilde K$ and $\tilde \mu$ are two complex constants which
depend on the electron density $n$. These complex constants are
expressed in terms of the more familiar viscosities, $\zeta$ (bulk
viscosity) and $\eta$ (shear viscosity) and elastic constants $K$
(bulk modulus) and $\mu$ (shear modulus) via the relations
\begin{eqnarray}
\tilde K(\omega)=K-i\omega \zeta,\\
\tilde \mu(\omega)=\mu-i\omega\eta,
\end{eqnarray}
where $\omega$ is the frequency of the external perturbation used to
excite the electron liquid.

The next step is then to {\it express} the visco-elastic
coefficients of the liquid in terms of its microscopic properties,
i.e., relate these quantities to the response functions. Here we
only report the results that are relevant to the present work and
refer the reader to Ref.~\cite{Conti1999} for an explicit derivation.
We are only concerned with the DC (zero frequency) limit of the
above quantities. By using an interpolation of the numerical results
of mode-mode coupling theory \cite{Nifosi1998} one finds
the following density dependence of the zero-frequency shear
viscosity (the bulk viscosity is identically zero in the same limit)
\cite{Conti1999}
\begin{equation}
\frac{\eta}{n}=\frac{1}{60 r_s^{-3/2}+80 r_s^{-1}-40
r_s^{-2/3}+62r_s^{-1/3}}
\label{eta3d}
\end{equation}
in 3D and in 2D by
\begin{eqnarray}
\left(\frac{\eta}{n}\right)^{-1}&=&\left(\frac{r_s^2}{12\pi}\ln\frac{2}{er_s}+0.25
r_s^2\right)^{-1}\nonumber \\
&&+21 r_s^{-2}+23 r_s^{-1/2}+13,
\label{eta2d}
\end{eqnarray}
where $r_s$ is the electron constant for the electron
liquid with uniform density $n$:
\begin{equation}
r_s a_B=\left\{
\begin{array}{cc} \left(3/4\pi n\right)^{1/3} & \mathrm{3D} \\
\left(1/\pi n\right)^{1/2} & \mathrm{2D}
\end{array} \right.,
\end{equation}
with $a_B$ the Bohr radius.

It is interesting to point out that specific confining potentials (e.g., an electron liquid in a quantum well) may make the approximations used to derive
Eqs.~(\ref{eta3d}) and (\ref{eta2d}) ill founded, leading to a
peculiar behavior of the viscosity coefficients \cite{DAgosta2007}.

\section{Hydrodynamic approach to transport in nanoscale systems}
\begin{figure}[ht!]
\includegraphics[clip,width=8cm]{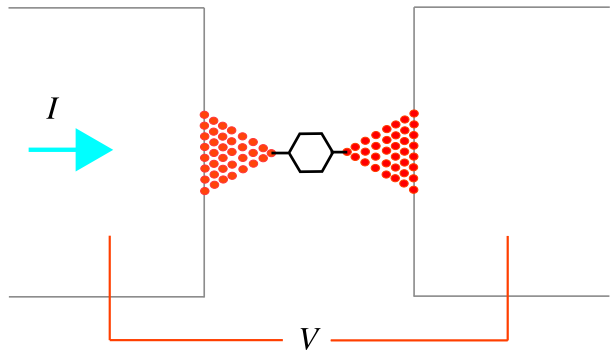}
\caption{Schematic of a nano-junction connected to two bulk
electrodes. A steady current is flowing from one electrode to the
other.} \label{nano-junction}
\end{figure}

In this section we show that in the case of nanoscale systems the
stress tensor can be approximated to a form similar to the classical
Navier-Stokes one. This is due to the geometric constriction
experienced by electrons flowing in the nanostructure which gives
rise to a very short ``collisional'' time~\cite{DiVentra2004a,
Bushong2005}. The system we have in mind is some nanoscopic junction
sandwiched between two mesoscopic or macroscopic leads (see Figure
\ref{nano-junction}) and current is induced in the system by, e.g.
polarizing the leads with a finite bias.
In this regime, we show that one can truncate the infinite hierarchy
of equations of motion for the electron stress tensor given in
(\ref{total-stress}) to second order and thus derive quantum
hydrodynamic equations. To realize how the simple presence of the
junction has such a strong impact on the equation of motion of the
current, one has to keep in mind that the former acts as a single
impurity potential that cannot be avoided by the electron flow. This
is different from the corresponding effect in bulk materials for
which a certain density of impurities is necessary to have a finite
resistance.

Let us then employ the quantum Boltzmann equation for the
single-particle distribution function $f(r,p,t)$ (which can be
derived from the time-dependent Schr\"odinger equation with standard
techniques ~\cite{Kadanoff}) and
show how the short collisional time induced by the nanostructure
allows us to close the equations for the stress tensor.
~\footnote{Clearly, for the definition of
local equilibrium distribution to be valid any length scale entering
the problem has to be larger than the system Fermi wavelength.}
The quantum Boltzmann equation for the distribution function in a co-moving
(Lagrangian) reference frame moving with velocity $v(r,t)$
is~\cite{DAgosta2006a,Tokatly1999}
\begin{eqnarray}
I[f]=&&D_{t}f(r,p,t)+\frac{p}{m}\nabla
f(r,p,t)+e\nabla \varphi \frac{\partial f(r,p,t)}{\partial p} -p\cdot
\nabla v\frac{\partial f(r,p,t)}{\partial p}\nonumber\\
&&-mD_{t}v\frac{\partial f(r,p,t)}{\partial p}
\label{kinetic}
\end{eqnarray}
where $I$ is the usual collision integral~\cite{Kadanoff}, $\varphi$
is the sum of the external potential and the Hartree part of the
interaction potential. The collision integral contains two terms,
one elastic and the other inelastic. In what follows, it is
important to realize that both terms can drive the system toward a
local equilibrium configuration.

From the quantum Boltzmann equation, we can derive the equation of
motion for the moments of the distribution function. The general
expression for the $m$th moment is the $m$th-rank tensor
\begin{equation}
P_{i_1,\ldots,i_m}=\int dp~p_{i_1}\ldots p_{i_m}f(p,r,t).
\end{equation}
The zeroth order is the single particle density, the first moment is
the velocity field, and the second moment is the stress tensor we
want to approximate. The equation of motion for the stress tensor
contains a term proportional to the third moment $P^{(3)}$:
\begin{eqnarray}
D_{t}P_{i,j}+P_{i,j}\nabla \cdot v+P_{i,k}\nabla _{k}v_{j}
+P_{k,j}\nabla_{k} v_{i}+\nabla_{k}P_{i,j,k}^{(3)}=\nonumber\\
\frac{1}{m}\int dp~ I[f]p_ip_j.
\label{secondmoment}
\end{eqnarray}
We note that $P^{(3)}$ enters in (\ref{secondmoment}) only through
its spatial derivative.  If the latter is small then the hierarchy
can be truncated \cite{DAgosta2006a,Tokatly1999}. From
(\ref{secondmoment}) we easily see that this derivative is small
compared to the other terms whenever
$\gamma=u/(L\max(\omega,\nu_{c})) \ll 1$. Here $u$ is the average
electron velocity, $L$ is the length of inhomogeneities of the
liquid that give rise to scattering among three particles, $\omega$
the system proper frequency and $\nu_{c}$ the collision rate.  The
parameter $1/L$ enters through the spatial derivative of $P^{(3)}$,
$\omega$ from the frequency dependence of the interactions (in the
DC limit of interest here $\omega\rightarrow 0$), $\nu_c$ through
the collisional integral $I[f]\propto -\nu_c(f-f_0)$, where $f_0$ is
the equilibrium Fermi distribution. This derivative is indeed small
for transport in nanostructures: When electrons move into a
nano-junction they adapt to the given junction geometry at a fast
rate, and approach to local equilibrium occurs at this fast rate
even in the absence of electron interactions~\cite{DiVentra2004a,
Bushong2005}. This ``relaxation'' mechanism occurs roughly at a rate
$\nu_c= (\Delta t)^{-1}\sim (\hbar /\Delta E)^{-1}$, where $\Delta
E$ is the typical energy spacing of lateral modes in the junction.
For a nano-junction of width $\ell$ we have $\Delta E\sim
\pi^{2}\hbar^{2}/m \ell^{2}$ and $\Delta t \sim m
\ell^{2}/\pi^{2}\hbar$. If $\ell=1~\mathrm{nm}$, $\nu_c$ is of the order of
$10^{15}$ Hz, i.e., orders of magnitude faster than typical
electron-electron or electron-phonon scattering rates. The condition
$\gamma=u/(L\max(\omega,\nu_{c}))\ll 1$ thus requires the length of
inhomogeneities $L\gg 1~\mathrm{nm}$, which is easily satisfied in
nanostructures. Note instead that in mesoscopic structures this
condition is not necessarily valid. In that case, the dominant
relaxation rate $\nu_c$ is given by inelastic effects, i.e. it is of
the order of THz, so that for typical lengths of mesoscopic systems,
$\gamma \approx 1$ in the DC limit. Nonetheless, the above condition
could still be valid for high-frequency excitations, like plasmons,
and/or very low densities, so that moments of the distribution of
order higher than two are negligible.

By neglecting $\nabla_{k}P_{i,j,k}^{(3)}$ in (\ref{secondmoment}) we
can thus derive a form for $P_{i,j}$. Let us write quite generally
the stress tensor $P_{i,j}$ as $P_{i,j}=\delta_{i,j}P-\pi_{i,j}$,
where the diagonal part gives the pressure of the liquid, and
$\pi_{i,j}$ is a traceless tensor that describes the shear effect on
the liquid. From (\ref{secondmoment}) we thus find that the tensor
$\pi_{i,j}$ can be written as (in $d$ dimensions, $d>1$)
\begin{equation}
\pi_{i,j}=\eta
\left(\nabla_{i}v_{j}+\nabla_{j}v_{i}-\frac{2}{d}\delta_{i,j}\nabla_{k}v_{k}\right)
\label{pixcstatic}
\end{equation}
where $\eta$ is a real coefficient (the viscosity) that is a
functional of the density~\cite{Tokatly1999}.  We point out that
(\ref{pixcstatic}) is in fact a particular case of a general stress
tensor with memory effects taken into
account~\cite{Conti1999,Vignale1996, Tokatly2005b}. In our
derivation this is the first non-trivial term of an expansion of the
stress tensor in terms of the density and velocity field.
Consequently the Navier-Stokes stress tensor in (\ref{pixcstatic})
can be seen as the first-order (non-trivial) contribution to the
exact stress tensor of the electron liquid (see also
~\cite{Tokatly2005a, Tokatly2005b, Vignale1996}).

Using this stress tensor we finally get from (\ref{newton-mb}) the
generalized Navier-Stokes equations for the electron liquid in
nanoscale systems
\begin{eqnarray}
D_{t}n(r,t)&=&-n(r,t)\nabla \cdot v(r,t),\nonumber\\
mn(r,t)D_{t}v_{i}(r,t)&=&-\nabla_{i}P(r,t)+\nabla_{j}\pi_{i,j}(r,t)\\
&&-n(r,t)\nabla_{i}V_{ext}(r,t).\nonumber
\label{completeNS}
\end{eqnarray}
Equations~(\ref{completeNS}) are formally equivalent to their
classical counterpart~\cite{Landau6} [see Eq.~(\ref{navier-stokes})]
and thus describe also nonlinear solutions, i.e., the possibility to
develop turbulence in the electron liquid in its normal state.  In
the following, we will consider only the case in which the liquid is
in the laminar regime and incompressible so that the viscoelastic
coefficients are spatially uniform. This latter approximation is
practically satisfied in metallic quantum point contacts (QPCs) but
needs to be relaxed in the case of QPCs with organic/metallic
interfaces (see, e.g., \cite{Sai2005}). In addition, for this case
the Hartree potential is constant and its spatial derivative is thus
zero.  Therefore, (\ref{completeNS}) reduce to the Navier-Stokes
equations for the density and velocity of a viscous but
incompressible electron liquid
\begin{eqnarray}
&&D_{t} n(r,t)=0,\nonumber\\
\label{NS} &&\nabla\cdot v(r,t)=0,\\
&&mn(r,t)D_{t}v_{i}(r,t)=-\nabla_{i} P(r,t)+\eta
\nabla^{2}v_{i}(r,t)-n(r,t)\nabla_{i}V_{ext}(r,t).\nonumber
\end{eqnarray}

\section{Heat equations from hydrodynamics}
The above results allow us to treat heat generation and transport
using a simplified hydrodynamic approach. In fact we know that the
flow of a viscous fluid, as described by our formalism, generates
internal friction and consequently an effective temperature
distribution inside the system. Therefore, when a steady state has
been reached, we can supplement the Navier-Stokes equations with an
equation for the energy balance. In the process of heat production,
we need to identify a heat source, a mechanism for the dissipation
of this heat and, since the system is in a steady state, equate
these two terms with the local entropy production. In a recent paper
\cite{DAgosta2006c} we have developed this model obtaining the
equation for the energy balance
\begin{equation}
\pi_{i,j}(r)\partial_j
v_i(r)+\nabla\cdot [k(r) \nabla T_e(r)]=c_{V}(T_{e})v(r)\cdot \nabla T_e(r),
\label{heatequilibrium}
\end{equation}
where $T_e$ is the electronic temperature, $k(r)$ is the diffusion
constant and $c_V$ is the specific heat at fixed volume of the
electron gas. \footnote{For an electron gas at low temperature,
$c_V=c_P$ since the correction is second order in temperature.}
Eq.~(\ref{heatequilibrium}) can be either justified on physical
grounds, or derived formally as high-order expansion of the
many-particle stress tensor \cite{Tokatly2005a}. We also stress once
more that in deriving this equation we have assumed that the flow of
the electron liquid is laminar, i.e., we are far from the onset of a
turbulent regime~\cite{Landau6,DAgosta2006a}. Obviously, in writing
Eq.~(\ref{heatequilibrium}) we have assumed that some thermodynamic
quantities like temperature and entropy for an electron liquid
flowing in a nanostructure can be defined. This is a much debated
point, and obviously we do not have a general solution for it.
However, here we argue that the electron temperature may be defined
as the one ideally measured by a probe weakly coupled to the system
and in local equilibrium with the latter \cite{DiVentra2008}. While
this operational definition may not be simple to realize in
practice, we know from experiments that local heat generation due to
current has a large effect on the stability of nanostructures
\cite{Zhifeng2007}.

From the form of Eq.~(\ref{heatequilibrium}) we can deduce a general
relation between the applied bias and the electron temperature. To
do this, we realize  \cite{DAgosta2006c} that the electron fluid
velocity, $v$, (which is generally smaller than the Fermi
velocity~\cite{Sai2007}) responsible for the transport of current
and heat is, in linear response, proportional to the bias $V$.
\footnote{$V$ may be given by an external battery, or the potential
due to a charge imbalance.} This simple proportionality, and the
usual result that $k\propto c_V$, bring us to the general result
\begin{equation}
T_e=\gamma_{ee} V,
\label{tedv}
\end{equation}
where $\gamma_{ee}$ is a constant whose expression in terms of
microscopic parameters of the electron liquid has been recently
derived for a quasi-adiabatic connection between the leads and the
nanojunction \cite{DAgosta2006c}
\begin{equation}
\gamma_{ee}=1.16\times\left(\frac{G}{nA_c}\right)\sqrt{\frac{d-1}{3d}\frac{\eta}{\gamma}}
\label{gammaee}
\end{equation}
where $G$ is the conductance of the system in the limit of zero
bias, $A_c$ its cross section, $d$ is the dimensionality ($d>1$).
Moreover,
\begin{equation}
\gamma=k_F^2 k_B^2 \lambda_{e}/9
\label{gamma3d}
\end{equation}
in 3D, and $\gamma=\pi k_{F}k_{B}^{2}\lambda_{e}/6$ in 2D \cite{DAgosta2006c}, $k_F$ is the Fermi
momentum, $k_B$ the Boltzmann constant, and $\lambda_e$ is the
inelastic mean free path.

Interestingly, Eq.~(\ref{tedv}) can be obtained from general
thermodynamic arguments, by comparing the energy dissipated in the
transport process in the nanostructure (proportional to $V^2$ from
Ohm's law), and the energy carried away by electrons (proportional
to $T_e^2$ for small temperatures)~\cite{DAgosta2006c}.

\section{Local electron heating}

In the case of a finite background temperature and in the absence of
ionic heating, from our hydrodynamic theory, the local temperature
of the electrons in the nanostructure is given by
\cite{DAgosta2006c}
\begin{equation} T_e(V)=\sqrt{T_0^2+\gamma_{ee}^2V^2}
\label{T1}
\end{equation}
where $V$ is the external bias, and $T_0$ is the electron
temperature deep into the electrodes.
If we now let the ions heat up, their effective local temperature is given by \cite{DAgosta2006c} (for values of the parameters
such that the argument in the root is non-negative)
\begin{equation}
T=\left(T_0^4+\gamma_{ep}^4
V^2-\gamma_{ee}^4V^4\right)^{1/4}\label{locheattemp}
\end{equation}
where $\gamma_{ep}$ can be expressed in terms of the physical
parameters of the nanostructure \cite{Todorov1998,Chen2005a}, and we
have assumed that both the ions and the electrons are at the same
background temperature $T_0$ deep into the electrodes. At zero
background temperature and for negligible electron-electron
interactions from the above equation we obtain the known result for the local
ionic temperature \cite{Todorov1998,Chen2005a,DAgosta2006c}
\begin{equation}
T\simeq \gamma_{ep}\sqrt{V}.
\end{equation}

{\em Effect on conductance --} We can now calculate the effect of
local electron heating on the conductance of a nanostructure. We
focus on the quasi-ballistic regime and we generalize Eq. (13) of
Ref. \cite{Chen2005a} for the inelastic current in the presence of a
finite electron temperature. \footnote{Note that a factor 2 is
missing in Eqs. (7) and (8) of Ref.~\cite{Chen2005a}.} We also
consider one mode frequency $\omega$. We will generalize later to
more modes. To take into account the effect of an effective local
electron temperature on the inelastic current, one faces the
calculation of terms with factors of the type $\int_{-\infty}^\infty
dE~f_E^{\alpha}(1-f_{E\pm\omega}^{\beta})$, with $\alpha,~\beta=\{R,L\}$
corresponding to electrons moving from either left or right, and
$f_E^{\alpha}=(\exp((E-\mu_{\alpha})/k_BT)+1)^{-1}$ is the Fermi
distribution with the difference between the electrochemical
potentials equal to the bias, $\mu_L-\mu_R=V$. (Refer to
\cite{Chen2005a} for additional details on the notation.)

We could provide a numerical calculation of the inelastic current.
However, we are interested in an analytical expression and thus
proceed as follows. We evaluate the above integrals in the
Sommerfeld approximation and keep only the terms of zeroth order in
the electron temperature (this is reasonable because the local
electron temperature is generally a small quantity). This
approximation brings us to the expression for the current flowing in
the system
\begin{equation}
I\simeq G_{el} V-G_{el} \frac{k_B}{\omega}\gamma_I
\left(T_0^4+\gamma_{ep}^4V^2-\gamma_{ee}^4
V^4\right)^{1/4}\frac{\log(e^{\beta(V-\omega)}+1)}{\beta},
\label{total_current}
\end{equation}
where $\gamma_I$ is the amplitude of the conductance drop at
$V=V_c\equiv \omega$ for zero electron and phonon temperature,
$G_{el}$ is the elastic conductance at zero bias, and
$\beta(V)=1/k_B T_e(V)$ where $k_B$ is the Boltzmann constant. By
differentiating Eq.~(\ref{total_current}) with respect to bias, and
again keeping only the terms of zeroth order with respect the
electron temperature, we arrive at
\begin{equation}
\frac{G}{G_{el}}\simeq 1 -\frac{k_{B}}{\omega }\gamma _{I}\left
(T_0^4+\gamma_{ep}^4V^2-\gamma_{ee}^4V^4\right )^{1/4}
\left[\frac{1}{e^{\beta(V)( \omega -V)}+1}\right].
\label{GG0}
\end{equation}
To obtain this result, one also has to bear in mind that the
approximations we make pertain to the energy region $V\simeq
\omega$, thus $d(\beta(V)(V-\omega)/dV\simeq \beta(V).$ An
expression for the conductance similar to Eq.~(\ref{GG0}) can be
derived for the case of zero electron temperature \cite{Chen2005a},
i.e., $\beta\to \infty$. Notice, however, that for consistency, one
has to take this limit in the expression for the current
(\ref{total_current}) {\it before} taking the derivative with
respect to the bias.

\begin{figure}[ht!]
\includegraphics[width=8cm,clip]{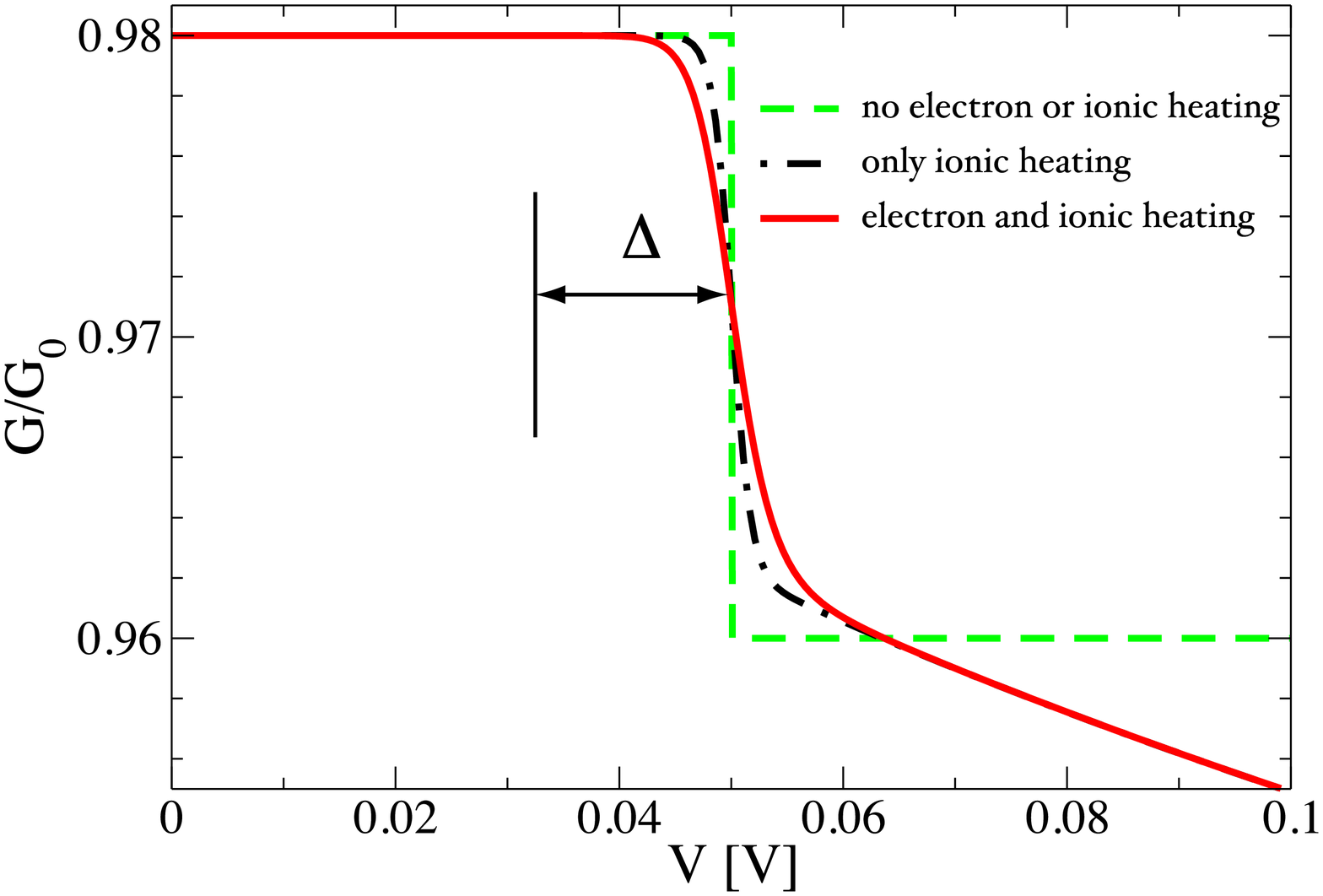}
\caption{Plot of $G/G_0$ as given by Eq.~(\ref{GG0}) where
$G_0=2e^2/h$. The  solid (red) curve has been generated with the following parameters:
$\gamma_{ee}=180~\mathrm{K/V},~\gamma_{ep}=600~\mathrm{K/\sqrt{V}},~G_{el}=0.98\times
G_0,~T_0=5~\mathrm{K}~, \gamma_I=0.02,~\hbar
\omega=0.05~\mathrm{eV}$. The dashed (green) line corresponds to the
case of zero phonon and electron temperature
($\gamma_{ee}=\gamma_{ep}=0$). The dashed-point (black) curve
corresponds to the case when only the phonon heating is taken into
account ($\gamma_{ee}=0$). In the figure we also define the parameter
$\Delta$ (see text).}
\label{theory}
\end{figure}

An example of the effect of local ionic and electron heating on
conductance is given in Fig.~\ref{theory} (see also discussion
below). In the absence of both effects (and at zero nominal
background temperature) the conductance shows a simple step-like
drop at the bias corresponding to the energy of the phonon mode. The
ionic heating introduces a shoulder at biases larger than the mode
energy, while the electron heating broadens the conductance curve
with an effective temperature larger than the nominal background
temperature.

{\em Comparison with experiments --} To compare our results with
available experimental data we consider a $\mathrm{D}_2$ molecule
sandwiched between two $\mathrm{Pt}$ leads \cite{Djukic2005}. We
focus on the predictions of our hydrodynamic theory on the local
electron heating effect. Therefore, we do not attempt to do a full
first-principles calculation of ionic heating, and take the relevant
parameters from experiment. For the $\mathrm{D}_2$ molecule we
consider a cross section of $\pi\times 1$~\AA$^2$, i.e., a circle
with radius $\simeq 1$~\AA. The nominal electron temperature deep
inside the electrodes is taken to be $T_0=5~\mathrm{K}$. From the
experimental results we have the frequency of the phonon mode
$V_c=0.05~\mathrm{eV}$, the drop of the conductance
\begin{equation}
\gamma_I=\left.\frac{\frac{d(I-I_0)}{dV}}{\frac{dI_0}{dV}}\right|_{V_c}=0.02,
\end{equation}
and the conductance at zero bias, $G_{el}=0.984~G_0$. We use as fit
parameters $\gamma_{ep}$ and evaluate $\gamma_{ee}$ from
Eq.~(\ref{gammaee}). In obtaining $\gamma_{ee}$ we have assumed an
inelastic mean free path $\lambda_e$ of $1~\mathrm{\mu m}$, a value
in line with the expectations for this system \cite{Pothier1997}. We
have also assumed that the electron density that enters the local
heating is the one of the chemical bonds between the $\mathrm{D}$
and Pt atoms. This density is estimated to be close to the Pt bulk
density, $n=6.6\times 10^{28}~\mathrm{m^{-3}}$ which gives the
electron constant $r_s\simeq 3$. From these values, the electron
viscosity $\eta$ and the constant $\gamma_{ee}$ are easily obtained
from Eqs.~(\ref{eta3d}) and (\ref{gamma3d}), respectively: The
electronic heating constant is predicted from Eq.~(\ref{gammaee}) to
be $\gamma_{ee}=180~\mathrm{K/V}$. This implies an effective
electron temperature of about 10 K at the $\mathrm{D}_2$ junction at
a bias of $50~\mathrm{mV}$. This temperature is higher that the
nominal bulk temperature.

The ionic heating constant is found to be $\gamma_{ep}=405$
K/$\sqrt{V}$. \footnote{This confirms that the ions heat more than
the electrons at the same bias and our approximation that leads to
Eq.~(\ref{locheattemp}) is justified.} This value can be compared
with the corresponding $\gamma_{ep}$ for a Au point contact at small
biases which is about $\gamma^{Au}_{ep}=170$ K/$\sqrt{V}$
\cite{Chen2005a}. This means that the Pt-$\mathrm{D}_2$-Pt system
heats up more than the Au QPC. For instance, at
0.1 V the ions of the Pt-$\mathrm{D}_2$-Pt junction have an average
temperature of about 130 K while at the same bias the gold atoms
heat up locally to about 54 K. This larger temperature is reasonable
since, while the conductance is similar for a Au point contact and
Pt-$\mathrm{D}_2$-Pt, the $\mathrm{D}_2$ molecule is lighter than Au
with a consequent increase of the electron-phonon coupling. In
addition, the modes of the $\mathrm{D}_2$ molecule have lower
probability to elastically scatter into the bulk modes of Pt -- thus
reducing lattice heat dissipation into the bulk electrodes -- than
the modes of a single Au atom into the bulk modes of Au. Both
effects lead to a higher local ionic temperature. We thus expect the
Pt-$\mathrm{D}_2$-Pt junction to be more unstable under the same
bias conditions than a Au point contact, i.e., we expect that the
chain Pt-$\mathrm{D}_2$-Pt breaks, on average, at much smaller
biases than Au point contacts due to heating effects.

\begin{figure}[ht!]
\includegraphics[width=8cm, clip]{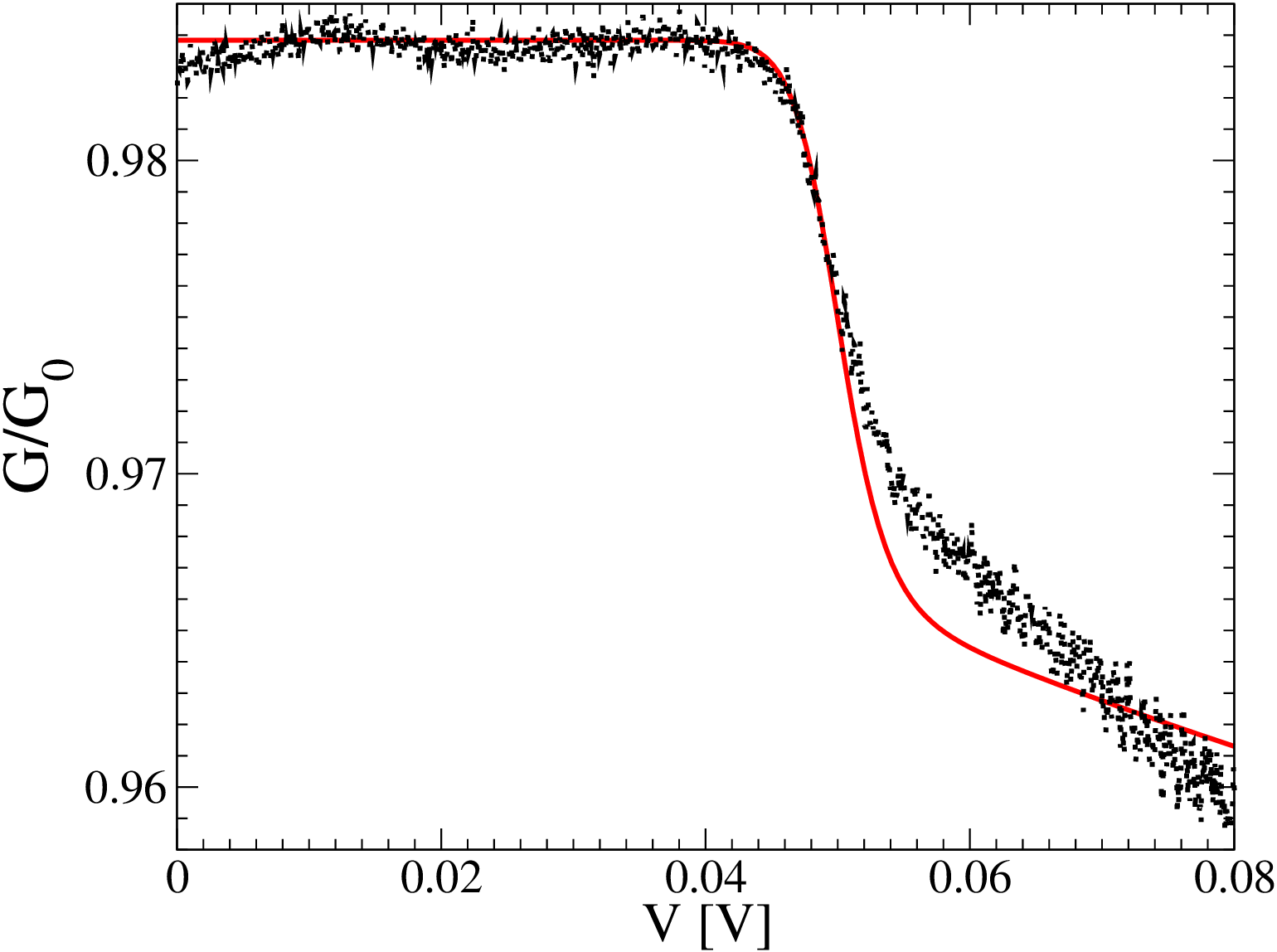}
\caption{Comparison between the experimental data \cite{Djukic2005} and
our theory [solid line, (red)] \cite{DAgosta2006a}.}
\label{experiment}
\end{figure}

The theoretical conductance containing both the local electron and
ionic heating effects is reported in Fig. \ref{experiment}, together
with the experimental data. The qualitative agreement between theory
and experiment is very good. It is interesting to note that the tail
of the experimental data goes approximately as $V^{0.8}$, while the
theory predicts $1-G/G_0\simeq V^{1/2}$ \cite{Todorov1998,Chen2003}. It
is important to realize, however, that at large biases, other
effects such as current-induced forces and other structural
instabilities may also contribute to the actual value of the
conductance \cite{Yang2005}.

We have also performed a second fit, not shown here, using
$\gamma_{ee}$ and $\gamma_{ep}$ as free parameters. The values for
these parameters obtained from this second fit are close to
those obtained from the theory and the one-parameter fit by less
than $10~\%$ (we find the best fit for
$\gamma_{ee}=200~\mathrm{K/V}$).

\begin{figure}[t!]
\includegraphics[width=8cm,clip,angle=0]{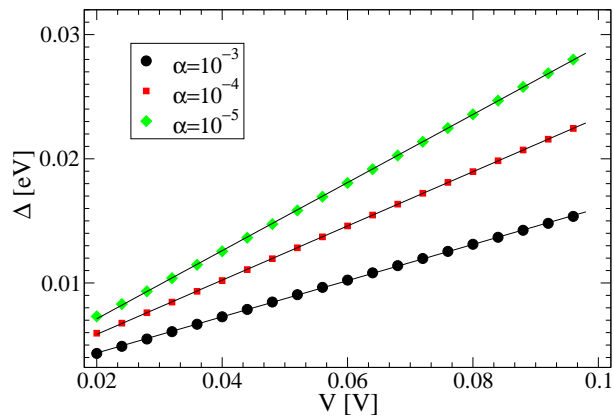}
\caption{$\Delta$ as a function of vibrational mode energy as defined
in the text. We have tested the linear behavior for different values
of the percentage $\alpha$ at which we calculate $\Delta$. The lines
that connect the symbols are linear regressions.}
\label{fig2}
\end{figure}

{\em Inelastic conductance width --} Let us now discuss how the width of
the inelastic conductance around the vibrational mode increases with
bias (see Fig.~\ref{theory}). This quantity can be directly measured and provides additional information on local electron heating.
If the background temperature is zero, the local electron
temperature increases linearly with bias as in Eq.~(\ref{T1}). Let
us define the quantity $\Delta$ as shown in Fig.~\ref{theory}: It is
the energy distance between the middle drop of the conductance and
the value at which the conductance assumes its purely elastic value
within a ratio $\alpha=[G_{el}-G(V_c)]/G_{el}$ as indicated in
Fig.~\ref{theory}. This quantity is plotted in Fig.~\ref{fig2} for
different values of the vibrational mode energy and for a few values
of $\alpha$, assuming that the vibrational energy is the only
quantity allowed to vary. We conclude that the width $\Delta$
increases linearly with the vibrational energy to reflect the linear
bias dependence of the local electronic temperature. A systematic
experimental study of this quantity would thus provide more
information on the electron heating phenomenon.

\section{Discussion} Our analysis in conjunction with the experimental data suggests that electrons heat up
locally at the $\mathrm{Pt-D_2-Pt}$ junction. Our Eq. (\ref{T1}) also predicts that electrons cool down when lowering the bias. On the other
hand, a {\em constant} electron temperature -- above the background
temperature -- for all biases is difficult to understand on physical
grounds, unless one assumes the existence of an external source of
energy that keeps the electron hot even at zero bias.

Experimental data showing an electron temperature equal to the
background temperature, i.e., negligible electron heating, may be
consistent with the fact that the effective cross section ``seen''
by the electron liquid is the one of a Pt atom and not of a
$\mathrm{D}_2$ molecule. \footnote{This system can be thought of as
a junction Pt-Pt-Pt with just one or few platinum atoms forming an
effective QPC which is not significantly affected
by the presence of the deuterium molecule.}  If that were the case,
the effective cross section would be 7 times larger than that of the
$\mathrm{D_2}$ molecule, and since the electron temperature scales
inversely proportional to the cross section (see Eq.~(\ref{gammaee})
and Ref. \cite{DAgosta2006a}) the electron heating temperature would
be lower than the background temperature. The conductance on the
other hand is unlikely to be so sensitive to this cross section due
to the extended nature of the Pt {\em d}-orbitals.

\begin{figure}[t!]
\includegraphics[width=8cm,clip]{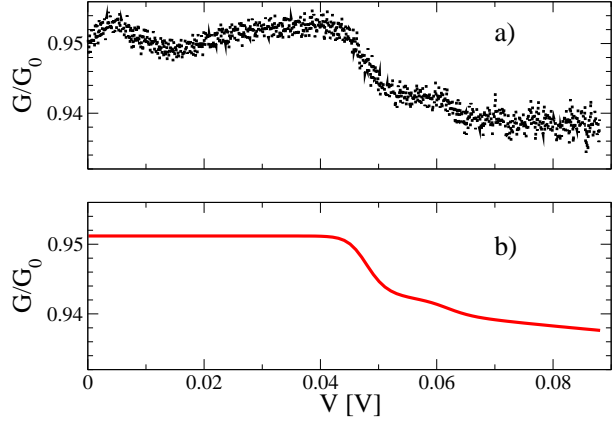}
\caption{a) Experimental results of the non-linear DC conductance of
a $\mathrm{H}_2$ molecule between two electrodes. The two steps
coming from the two phonon modes at $48$ and $62~\mathrm{meV}$ are
clearly visible, together with some unexpected structure at low
bias. b) Plot of $G_{H_2}$ as a function of the external bias
according to Eq.~\ref{twomodes}. In this plot, that is not a fit to
the experimental data, we have used $\gamma_{I1}=0.01$,
$\gamma_{I2}=0.002$, $\gamma_{ee}=180~\mathrm{K/V}$,
$\gamma_{ep,1}=\gamma_{ep,2}=400~\mathrm{K}/\sqrt{\mathrm{V}}$. The
other parameters are the same as in Fig.~\ref{experiment}.}
\label{Gh2}
\end{figure}

Further generalization of Eq. (\ref{GG0}) to the case where many
vibrational modes are present is possible. For example, it has been
reported that a $\mathrm{H_2}$ molecule sandwiched between two Pt
leads shows two fundamental vibrational frequencies at
$48~\mathrm{meV}$ and $62~\mathrm{meV}$ \cite{Djukic2005}.
If one assumes that scattering by these two modes is uncorrelated, a straightforward generalization of Eq. (\ref{GG0}) leads to
\begin{eqnarray}
\frac{G_{H_2}}{G_{el}}\simeq&&1 -\frac{k_{B}}{\omega_1 }\gamma
_{I1}\left (T_0^4+\gamma_{ep,1}^4V^2-\gamma_{ee}^4V^4\right )^{1/4}
\left[\frac{1}{e^{\beta(V)( \omega_1 -V)}+1}\right]\nonumber\\
&&-\frac{k_{B}}{\omega_2 }\gamma _{I2}\left
(T_0^4+\gamma_{ep,2}^4V^2-\gamma_{ee}^4V^4\right )^{1/4}
\left[\frac{1}{e^{\beta(V)( \omega_2
-V)}+1}\right],
\label{twomodes}
\end{eqnarray}
where $\omega_1$ and $\omega_2$ are the two vibrational frequencies
and we have taken into account the possibility that the two coupling
constants $\gamma_{ep,1}$ and $\gamma_{ep,2}$, and the two
amplitudes of the conductance drops $\gamma_{I1}$ and $\gamma_{I2}$
be different.  A plot of $G_{H_2}$ is reported in Fig. \ref{Gh2} as
a function of the external bias along with the experimental data.
Since the cross section for $\mathrm{D}_2$ and $\mathrm{H}_2$ is essentially the same,
and the electron heating does not depend on the mass of the ions,
our estimate of $\gamma_{ee}$ holds for $\mathrm{H}_2$ as well.
Our results are again in qualitative agreement with the available
experimental data \cite{Djukic2005}, although our theory might be
not sufficient to quantitatively describe all the experimental
findings. Indeed, our fit in this case has failed in producing any
sensible result for the constants $\gamma_{ee}$, $\gamma_{ep,1}$ and
$\gamma_{ep,2}$: the large fluctuations of the experimental data,
especially in the region of small bias and close to the phonon modes
energies do not allow for a systematic fit of the data with the
theory. Finally, it is interesting to note that a value of
$\gamma_{ep}$ similar to the one we have obtained for the
$\mathrm{D_2}$ molecule gives a reasonably good agreement between
theory and experiment also for the $\mathrm{H_2}$ molecule. This
seems to suggest that the longitudinal modes of the bonds between
the $\mathrm{H}$ and $\mathrm{Pt}$ atoms are mainly responsible for
the local ionic heating of the Pt-$\mathrm{H}_2$-Pt junction, and
similarly the longitudinal modes of the bonds between the
$\mathrm{D}$ and $\mathrm{Pt}$ for the Pt-$\mathrm{D}_2$-Pt
junction. We expect that such modes are slightly affected by the
change of mass of the smaller atom in the bond. Clearly, more
theoretical and experimental work in this direction is necessary.

\section{Conclusions} We have discussed a novel hydrodynamic approach to transport that allows the description of charge and heat flow in terms of the single-particle density and velocity field of the electron liquid \cite{DAgosta2006a}. The theory allows us to make predictions about the electron
flow past a nanostructure and its dependence on the external bias
(or the current). One such prediction is the heating of electrons
locally at the nano-junction \cite{DAgosta2006c}. Here we have
considered the measurable consequences of this effect on the
inelastic conductance which shows a broadening at the inelastic step
larger than the one expected from the background nominal
temperature. We have compared our theory with available experimental
results \cite{Djukic2005} and found a reasonable quantitative
agreement for the case of a $\mathrm{D_2}$ molecule between two Pt
leads.  For the case of a $\mathrm{H_2}$ molecule between the same
leads our theory is only in qualitative agreement with the
experimental findings. We also predict that the width of
the inelastic conductance step should increase linearly with bias, a
fact that can be tested experimentally.

\section*{Acknowledgments} We acknowledge financial support from the
Department of Energy grant DE-FG02-05ER46204. We thank the authors
of Ref.~\cite{Djukic2005} for the use of their data and useful
discussions.

\section*{References} \bibliographystyle{iopart-num}
\bibliography{mine,books,articles}

\end{document}